\documentclass{article}
\usepackage{graphicx}			
\usepackage{booktabs}			
\usepackage{tabulary}			
\usepackage[T1]{fontenc}		
\usepackage[utf8]{inputenc}		
\usepackage{xpatch}
\usepackage{xcolor}				
\usepackage{listings}			
\usepackage[page,toc,title]{appendix}
\usepackage{minitoc}
\usepackage{enumitem}
\usepackage{hyperref}
\hypersetup{
   colorlinks=true,
   linkcolor=black,
   citecolor=black,
   filecolor=magenta
}
\usepackage{apacite}

\usepackage{multibib}
\newcites{SI}{Supplementary Information References}

\usepackage{authblk}
\usepackage{float}
\usepackage{tabularx,ragged2e}
\usepackage{longtable}
\usepackage{array}

\newcommand{\SIref}[1]{Supplementary Information~\ref{#1}}

\newcolumntype{L}[1]{>{\raggedright\let\newline\\\arraybackslash\hspace{0pt}}m{#1}}

\title{\vspace{-4cm}Artificial Intelligence in Government: Why People Feel They Lose Control}
\author[1]{Alexander Wuttke \thanks{a.wuttke@lmu.de}}
\author[2]{Adrian Rauchfleisch } 
\author[3]{Andreas Jungherr}

\affil[1]{Geschwister-Scholl-Institute of Political Science, LMU Munich}
\affil[2]{Graduate Institute of Journalism, National Taiwan University}
\affil[3]{Institute for Political Science, University of Bamberg}

\date{May 2025}

\begin{document}
\doparttoc 
\faketableofcontents 

\maketitle

\begin{center}
Working Paper
\end{center}

\begin{center}
\textbf{Abstract}
\end{center}
\begin{quote}
The use of Artificial Intelligence (AI) in public administration is expanding rapidly, moving from automating routine tasks to deploying generative and agentic systems that autonomously act on goals. While AI promises greater efficiency and responsiveness, its integration into government functions raises concerns about fairness, transparency, and accountability. This article applies principal-agent theory (PAT) to conceptualize AI adoption as a special case of delegation, highlighting three core tensions: assessability (can decisions be understood?), dependency (can the delegation be reversed?), and contestability (can decisions be challenged?). These structural challenges may lead to a “failure-by-success” dynamic, where early functional gains obscure long-term risks to democratic legitimacy. To test this framework, we conducted a pre-registered factorial survey experiment across tax, welfare, and law enforcement domains. Our findings show that although efficiency gains initially bolster trust, they simultaneously reduce citizens’ perceived control. When the structural risks come to the foreground, institutional trust and perceived control both drop sharply, suggesting that hidden costs of AI adoption significantly shape public attitudes. The study demonstrates that PAT offers a powerful lens for understanding the institutional and political implications of AI in government, emphasizing the need for policymakers to address delegation risks transparently to maintain public trust.

\end{quote}

\section{Introduction}
The deployment of Artificial Intelligence (AI) in public administration is accelerating \cite{margetts_how_2024}. Initially heralded for its ability to automate routine and rule-based tasks, recent advances in generative and agentic AI—systems capable not only of producing content but also of autonomously acting on goals—have expanded the range of public sector applications \cite{acharya_agentic_2025, park_generative_2023,purdy_what_2024}. These systems promise to enhance efficiency and responsiveness in service delivery, for example, by guiding citizens through complex bureaucratic processes or flagging inconsistencies in large-scale data systems \cite{straub_ai_2024}. Governments worldwide are exploring AI’s potential to streamline routine tasks and citizen interactions across domains such as tax processing, benefits eligibility, and law enforcement risk assessments.

While early deployments often focus on measurable efficiency gains and user satisfaction, deeper questions persist. These concern the fairness and accuracy of AI decisions \cite{barocas_fairness_2023,maleki_ai_2024}, the presence of data biases \cite{buyl_large_2024}, system opacity \cite{burrell_how_2016}, and the accountability of automated decision-making \cite{ananny_seeing_2018}. These tensions, though often latent in early phases of AI adoption, are likely to shape long-term public evaluations of AI in government. How will artificial intelligence influence the relationship between the state and its citizens? 

This article employs principal-agent theory to conceptualize the delegation of governmental authority to AI systems, revealing foreseeable tensions inherent in this process. Principal-agent theory (PAT) provides a well-established framework for analyzing institutional arrangements in which one actor (the principal) delegates tasks to another (the agent) under conditions of uncertainty, asymmetric information, and limited control \cite{miller_political_2005}. From the perspective of PAT, the principal-agent relationship is defined by a control problem: principals lack either the competence or the information necessary to fully oversee the agent’s decisions.

We argue that delegating tasks to AI systems represents a distinct variant of the principal-agent problem, giving rise to the fundamental challenge of maintaining control. Despite lacking intentionality, AI systems act on behalf of principals and produce outputs with real-world consequences for citizens and state institutions. Conceptualizing AI as an agent within a PAT framework enables us to build on established theoretical insights to better understand the opportunities and risks associated with delegating authority to AI. As in classical principal-agent relationships, assigning tasks to AI in public administration introduces three central tensions:

\begin{itemize}
  \item \textbf{Assessability}: AI systems, particularly those built on complex machine learning models, are often opaque, making it difficult for principals to monitor and understand their decision-making processes.
  \item \textbf{Dependency}: Over time, reliance on AI can erode institutional expertise and create path dependencies that limit the reversibility of delegation.
  \item \textbf{Contestability}: AI systems may lack clear, accessible mechanisms for challenging decisions, thereby weakening traditional channels of administrative accountability.
\end{itemize}

We conceptualize these tensions as structural features of AI-enabled governance that may lead to a “failure by success” dynamic. That is, early functional gains in efficiency may obscure longer-term risks. Initially, the AI use in government may foster the legitimacy of a political system by delivering tangible benefits, which may lead to more widespread use of AI. But over time, the tensions inherent to delegating tasks to AI may become more apparent, leading to a diminished sense of control and democratic accountability. 

To understand these risks, we provide a theoretical framework derived from PAT that links functional delegation at the institutional level to political legitimacy at the mass level. To empirically test this framework, we field a factorial survey experiment in the United Kingdom using a vignette design across three domains: tax administration, welfare benefits, and bail risk assessments. Each vignette describes either human-led or AI-enabled service provision, with variations that highlight assessability, dependency, or contestability concerns. We pre-registered our design and hypotheses, expecting that while efficiency-focused AI deployments might raise trust, awareness of delegation risks will reduce perceived control and support for further AI use.

Indeed, we observe increased trust in government when highlighting tangible efficiency gains resulting from AI systems used for routine government tasks. However, even this favorable description of AI deployment simultaneously undermines citizens’ perceived control. This ambivalence reflects a fundamental trade-off: while people appreciate AI’s benefits, they also feel they have lost control over democratic decisions when AI is used in public administration. Even worse, when AI use is framed to highlight deeper principal-agent tensions—specifically, the lack of assessability, creation of irreversible dependencies, or absence of contestability—both institutional trust and perceived control decline sharply. These effects are robust across all tested policy domains and suggest that awareness of these delegation risks drives public opposition to expanded AI use in government. In line with principal-agent theory, our results demonstrate that even successful AI implementation may generate system-undermining attitudes, revealing a “failure-by-success” dynamic in democratic governance.

This study makes two theoretical contributions. First, it advances the literature on the societal implications of artificial intelligence by introducing principal-agent theory as a lens to analyze AI in public governance. Conceptualizing AI as a delegated agent, the framework developed here draws on PAT to identify key dimensions that shape AI's role in government.  Second, it contributes to the application of PAT in public administration by examining the well-known challenges of delegation and control from a novel perspective. While political scientists typically study control problems through the lens of principal-agent theory with a focus on institutional arrangements, this research shifts attention to the psychological effects experienced by individual citizens when AI exercises power and citizens feel a loss of control over their own destinies \cite{averill_personal_1973, bandura_self-efficacy_1996}—an outcome with potential implications for democratic legitimacy. 

Our findings suggest that principal-agent theory not only helps explain the institutional dynamics underlying AI adoption in government but also offers predictive insight into how citizens may respond to the less visible costs of automated governance. These results underscore the importance for policymakers to recognize the inherent tensions in delegating public tasks to AI systems and to communicate transparently about how these tensions are being addressed. Failing to do so risks not only eroding public support for AI use, but also diminishing trust in government more broadly.

\section{Principal Agent Theory for AI}
\subsection{AI in public administration}
Recent advances in the development and deployment of generative AI—stemming from foundational work such as \citeA{vaswani_attention_2017} on the Transformer architecture and subsequent breakthroughs like GPT-3 \cite{brown_language_2020}—have drawn widespread attention to the creative and expressive capacities of AI systems. This focus on generative capabilities, amplified by user-facing applications such as ChatGPT \cite{openai_introducing_2022}, has somewhat overshadowed earlier public discourse that emphasized AI's utility in automating routine and repetitive tasks \cite{agrawal_prediction_2018, autor_skill_2003}. However, the public and research attention is beginning to shift again as developments in so-called agentic AI gain traction—systems that integrate generative abilities with autonomous decision-making and action execution \cite{acharya_agentic_2025, park_generative_2023, purdy_what_2024}. These agentic systems, often embedded in interactive interfaces, can perform complex tasks on behalf of users, signaling a convergence between generative intelligence and goal-directed behavior.

Agentic AI is one type of AI-enabled system that increasingly is used in public administration. These systems are widely seen to hold great potential for application in public administration tasks \cite{committee_of_public_accounts_use_2025, european_commission_directorate_general_for_informatics_analysis_2025,margetts_rethink_2019, straub_ai_2024}. For example, an AI-powered virtual assistant could autonomously guide citizens through complex administrative processes\textemdash such as applying for housing subsidies\textemdash by interpreting user input, retrieving relevant data, generating personalized forms, and submitting applications on their behalf.

The same underlying logic can be translated to other tasks and contexts. One prominent area of application is the automation of routine government services such as passport renewals, driver’s license applications, and tax return processing. These services involve a vast number of micro-transactions\textemdash estimated at around one billion annually in the United Kingdom alone \cite{straub_ai_2024}\textemdash many of which follow standardized, rule-based procedures. As such, they are particularly well suited for delegation to agentic AI systems. These systems can perform functions like detecting errors in submitted forms, categorizing applications based on eligibility criteria, or flagging anomalies for review, thereby freeing human personnel to focus on more complex or discretionary cases. More advanced applications can even render preliminary decisions in routine scenarios while escalating ambiguous cases to human decision-makers. For citizens, this can result in shorter processing times, fewer administrative errors, and a more seamless experience—ultimately contributing to higher levels of trust and satisfaction with public services \cite{austin_generative_2024, downie_ai_2024, richter_navigating_2024, microsoft_transforming_2024}.

AI-enabled systems hold the promise of reducing bureaucratic friction, improving service accessibility, and enabling more efficient, user-centered delivery of public services. However, their adoption also raises significant concerns. These include questions about the alignment of AI systems with the intentions of both developers and end-users \cite{amodei_concrete_2016, bommasani_opportunities_2022, jungherr_what_2025}, the accuracy and fairness of algorithmic decisions \cite{barocas_fairness_2023, maleki_ai_2024, maynez_faithfulness_2020}, the presence of biases in training data \cite{buyl_large_2024}, the ethical implications of AI deployment \cite{hagendorff_ethics_2020}, and the accountability mechanisms in place when failures occur \cite{ananny_seeing_2018}. Such issues can shape how individuals experience and evaluate interactions with AI-enabled public services. While efficiency gains and faster service delivery may drive initial acceptance, these deeper concerns are likely to influence broader public attitudes toward AI in government over time.

Currently, direct experiences with AI-enabled government services are still rare. Assessing whether the underlying issues outlined above will significantly impact their future use and public perception is ultimately a prognostic task. Although the future cannot be predicted with certainty, theoretical frameworks can help us anticipate potential risks and dynamics. To assess whether early perceptions of gains in efficiency and cost-reduction through AI might give way to longer-term perceptions of governance failures, we use principal-agent theory to identify key areas of tension that emerge when public tasks are delegated from principals (e.g., government bodies) to agents (e.g., AI systems or contractors), including challenges of control, trust, and willingness to delegate.

\subsection{AI in public administration as a Principal-Agent Problem}
Principal-agent theory offers a powerful framework for analyzing the delegation of authority in modern governance. PAT focuses on situations in which a principal—such as an elected official, administrative supervisor, or the public—delegates tasks or decision-making power to an agent, who is expected to act on the principal’s behalf under conditions of uncertainty and imperfect control. A fundamental concern in this literature is that delegation introduces persistent coordination problems, including information asymmetries, preference divergence, and limited monitoring and sanctioning capacity \cite{miller_political_2005}.

While PAT has traditionally been used to analyze relationships between political actors—such as legislators, bureaucrats, and regulatory agencies \cite{downs_conflict_1994, gailmard_agency_2009, kiewiet_logic_1991, strom_delegation_2003}—its analytical reach can be extended to include increasingly technocratic forms of delegation. One such development is the incorporation of AI-enabled systems into public administration. When governments delegate routine or even discretionary tasks to AI these systems function as non-human agents whose outputs shape the material conditions of governance. Despite lacking consciousness or intentionality, we can model task delegation to AI systems along core assumptions as specified in canonical PAT formulations \cite{holmstrom_moral_1979, shavell_risk_1979, miller_political_2005}. According to Miller \citeA[p. 205f.]{miller_political_2005}, the canonical principal-agent model builds on six core assumptions. Each of these applies when examining the deployment of AI in government:

First, an agent’s action must meaningfully affect outcomes for the principal. AI systems clearly meet this criterion: whether reviewing tax returns, processing immigration applications, or flagging social welfare fraud, AI actions produce outputs that have direct effects on policy delivery, resource allocation, and citizens’ rights and entitlements \cite{eubanks_automating_2017}.

Second, the principal cannot directly observe the agent’s actions and must rely on outcomes to infer performance. In the case of AI, this information asymmetry is amplified. Most AI systems\textemdash particularly those using deep learning\textemdash are functionally opaque. Their decision processes are difficult to audit or interpret, even by technically trained administrators, making effective monitoring challenging and costly \cite{burrell_how_2016}.

Third, agents are assumed to be self-interested and thus may “shirk” if not adequately incentivized. AI systems lack intention, but outputs can diverge from the principal’s goals due to biases in training data, errors in model specification, or poorly defined objectives. This functional misalignment represents a novel form of preference asymmetry\textemdash technically induced rather than motivational\textemdash requiring oversight and correction mechanisms analogous to those used for human agents.

Fourth, delegation originates from a unified principal capable of structuring the terms of delegation. In democratic systems, this includes policymakers and senior administrators who decide whether and how to implement AI. Governments initiate contracts with AI vendors, define scopes of authority, and (in principle) determine the accountability structure governing AI use.

Also, principals and agents share common knowledge of the rules of the game and that agents are rational actors optimizing against constraints. With AI, rational optimization occurs within system parameters set during development. While AI agents are not conscious, they “rationally” follow algorithms designed to maximize specified objectives, often based on large-scale empirical learning from past data. Developers and public managers may understand the goal structure but not always the detailed behavior of models in novel contexts, complicating backward induction and control.

Finally, canonical models assume the principal holds ultimate bargaining power and can structure “take-it-or-leave-it” contracts. In public administration, delegation to AI often occurs under constraints imposed by external vendors, standardized tools, or infrastructural dependencies, reducing the principal’s effective control. Moreover, once AI replaces human expertise, the ability to renegotiate or revoke delegation becomes constrained by capability loss and sunk costs.

From this perspective, principal-agent theory offers a compelling foundation for analyzing the control dynamics involved in delegating authority to AI systems in government. When considering the conditions under which governments deploy AI\textemdash particularly as framed by the canonical assumptions of PAT\textemdash it becomes clear that this form of delegation aligns with what \cite{abbott_competencecontrol_2020} identify as a competence-control dilemma. Their work shows that increasing competence asymmetries between principals and agents can invert power relationships: while principals typically seek capable agents to carry out delegated tasks, agents whose expertise substantially exceeds that of their principals can become difficult to monitor or override.

Applying this insight to AI delegation in public administration reveals three key tensions that help us structure the evolving relationship between governments and AI systems. These tensions give rise to three interrelated challenges that shape the political and administrative consequences of algorithmic delegation:

\begin{itemize}
  \item \textbf{Assessability}: The complexity and opacity of AI systems make it difficult for principals to evaluate how decisions are made. This exacerbates existing monitoring problems by deepening information asymmetries and reducing the effectiveness of traditional oversight mechanisms.
  \item \textbf{Dependency}: As governments reduce human administrative capacity in favor of automated systems, delegation becomes less reversible. Over time, this creates structural lock-in, where the agent’s operational role cannot easily be curtailed or reconfigured, even in the face of poor performance or public opposition.
  \item \textbf{Contestability}: AI systems often lack clear, accessible avenues for contesting decisions, which undermines the mechanisms through which principals can sanction agent behavior or demand corrective action. Once authority is delegated, principals may find it difficult to reassert control or enforce normative standards.
\end{itemize}

Importantly, these challenges do not arise from the malfeasance or self-interest of AI agents—as in classical formulations of PAT—but from the technical complexity, structural opacity, and institutional entrenchment that characterize modern AI systems. Nonetheless, the implications are analogous: the more capable, autonomous, and opaque the agent becomes, the more difficult it is for the principal to exercise meaningful control.

We argue that these three challenges\textemdash assessability, dependency, and contestability\textemdash offer a productive framework for analyzing how governments delegate to AI, and for understanding public reactions to the growing use of AI in public administration.

Framing governmental use of AI as a specific instance of control delegation within principal-agent theory highlights how AI in public administration may exemplify a case of failure by success. As outlined above, various perspectives anticipate that AI will deliver efficiency and cost savings. These functional improvements could enhance state capacity, even under conditions of fiscal discipline, which would clearly represent a success. Indeed, early studies indicate that the public tends to favor AI applications under specific conditions \cite{araujo_ai_2020, ingrams_ai_2022, raviv_when_2023}. However, the tension between control and competence, as identified by Abbott \citeA{abbott_competencecontrol_2020}, suggests that initial success may paradoxically lead to failure when AI is used more broadly—failure through loss of control. We know from other political contexts that people often react negatively to uses of AI they feel are transgressive or beyond their control \cite{jungherr_deceptive_2024, jungherr_artificial_2025, rauchfleisch_explaining_2025}. A failure through success of governmental uses of AI could manifest materially as increased governmental dependency on AI systems, a trend that might be traced through policy research. Yet, given AI’s growing public visibility, such failure through success may also surface in shifting public attitudes toward both AI and governmental authority. This is the issue to which we now turn.

\subsection{AI in public administration: Failure by success}
The ultimate success of AI use in government depends on public willingness to accept and trust AI-enabled services. In the short term, positive experiences and narratives centered on efficiency gains and cost reduction are likely to foster supportive reactions. Over time, however, underlying tensions in the use of AI in public administration may generate a critical public response. As we are still in the early stages of AI adoption, direct evidence of negative reactions to widespread AI deployment remains limited. Nonetheless, the tensions identified through Principal-Agent Theory offer a framework for anticipating likely points of conflict that may shape public attitudes in the future.

Principal-agent theory highlights the central role of control in the delegation of tasks from principals to agents. In the context of AI in government, the public may be seen as the ultimate principal, facing the consequences of administrative decisions to delegate tasks to AI systems. Public encounters with such delegation may generate positive responses, such as improved efficiency and service quality, but also negative ones, especially when they evoke a perceived loss of control, declining trust in the delegating authority, or discomfort with the nature of the delegation itself.

Building on these concerns, the core dimensions identified through PAT, assessability, dependency, and contestability, suggest how the early success of AI in public administration might paradoxically contribute to its eventual failure, particularly in the court of public opinion.

\subsubsection{Assessability}
Assessability is a major concern in AI deployment and is reflected in the explainable AI (XAI) subfield \cite{barredo_arrieta_explainable_2020, dwivedi_explainable_2023, gunning_xaiexplainable_2019}, which develops methods to make black-box models interpretable and, therefore, more trustworthy. A central concern in PAT is the difficulty principals face in overseeing agents, especially when agents possess superior technical expertise or privileged access to information. This information asymmetry impairs the principal’s ability to ensure agents remain aligned with the principal’s objectives \cite{kiewiet_logic_1991, miller_political_2005}. Large-scale AI systems\textemdash especially those built on complex machine learning models\textemdash exacerbate this problem. Their “black box” nature renders their internal logic opaque, even to trained administrators \cite{burrell_how_2016}. This lack of transparency hinders effective oversight, undermines procedural accountability, and weakens confidence in the responsiveness of the agent. From a PAT perspective, low assessability disrupts the conditions under which delegation is rational and beneficial.

Public awareness of the opacity and complexity of AI systems in government is therefore likely to diminish willingness to delegate decision-making to AI, reinforce feelings of lost control, and potentially reduce trust in the government responsible for the delegation.

\noindent\textbf{Assessability Hypothesis}: The perception that AI systems are insufficiently assessable undermines public support and weakens the principal’s ability to monitor agent behavior.

\subsubsection{Dependency}
While principals generally prefer competent agents, excessive competence asymmetries can invert the delegation relationship, leaving the principal unable to effectively guide or revoke the agent’s actions \cite{abbott_competencecontrol_2020}. This risk is particularly salient in the context of AI. As governments automate administrative tasks, they may inadvertently erode internal human expertise. Over time, this produces organizational path dependencies \cite{pierson_increasing_2000} that make reversing or modifying AI-based decisions increasingly costly and impractical—even when performance falters or public values shift \cite{flinders_shrinking_2012}. What begins as a move toward efficiency can ultimately restrict institutional flexibility and undermine the possibility of reasserting human oversight. In PAT terms, the agent’s autonomy can surpass the principal’s capacity for retraction, creating structural imbalances.

If the public comes to view AI systems as creating lock-in effects or institutional overreliance, their support for such systems is likely to wane, accompanied by reduced willingness to delegate and growing concerns over loss of control and government accountability.

\noindent\textbf{Dependency Hypothesis}: The perception that AI systems create irreversible dependencies undermines public support and reflects diminished reversibility in delegation.

\subsubsection{Contestability}
Another key concern in PAT is the principal’s ability to sanction or override agent behavior that deviates from intended goals \cite{weingast_bureaucratic_1983}. In democratic governance, this ability is typically preserved through mechanisms for contestability, such as appeals, oversight bodies, or administrative redress. Yet, even where human oversight mechanisms are formally retained for tasks delegated to AI, practical constraints—such as automation bias or lack of resources—can render them ineffective \cite{narayanan_ai_2024}. As a result, delegation becomes both less reversible and less accountable—two outcomes that PAT identifies as threats to legitimacy.

When the public perceives that AI decisions cannot be meaningfully questioned or challenged, their support is likely to diminish. Such perceptions reinforce broader anxieties about lack of control, unaccountable governance, and democratic erosion.

The “failure by success” dynamic emerges when governments adopt AI to improve efficiency and service delivery but, in doing so, undermine key institutional principles of transparency, reversibility, and contestability. While AI may deliver measurable short-term benefits, its longer-term effects on public perceptions of trust, control, and accountability may ultimately jeopardize democratic legitimacy and the sustainability of administrative reform.

\noindent\textbf{Contestability Hypothesis}: The perception that AI systems are incontestable undermines support for their use and reflects a breakdown of accountability mechanisms.

\section{Research design}
We use a factorial survey experiment \cite{auspurg_factorial_2015} to test whether making people aware of issues related to assessability, dependency, and contestability affects people’s sense of control, trust in government, and willingness to delegate public administration tasks to AI. Respondents were shown vignettes describing how governmental tasks are currently performed or could be performed using AI. The vignettes systematically varied key dimensions based on our theoretical framework. This approach allows us to isolate and measure causal effects while maintaining high ecological validity through realistic scenario descriptions.

Each respondent evaluated three distinct vignettes presented sequentially. This is similar to a multiple-conjoint task design \cite{hainmueller_causal_2014}, which enhances statistical power by increasing the number of observations and allows comparison across different types of scenarios. Additionally, we vary AI use in three domains of government, which resembles a topic-sampling design \cite{clifford_estimators_2024}. The goal is to avoid over-reliance on any single context and to improve generalizability. The areas of application focus on three policy domains: tax administration, welfare benefits approval, and bail decision risk assessment. The presentation order of domains was randomized to mitigate potential sequence effects.

For each domain (tax, welfare, bail), respondents will be randomly assigned to one of four conditions that each resemble different approaches to AI uses (see \SIref{SI:vignettes} for the wording of the vignettes). 

\begin{itemize}
  \item \textbf{Human condition}: Describes how the task is currently performed by humans, serving as the reference category for all analyses.
  \item \textbf{AI Benefits Condition}: Outlines how AI could execute the task, emphasizing its positive effects.
  \item \textbf{AI Benefits + Dependency Condition}: Includes the AI Benefits Condition plus an additional paragraph highlighting the risks of excessive dependence on AI and the erosion of human state capacity.
    \item \textbf{AI Benefits + Contestability Condition}: Includes the AI Benefits Condition plus a paragraph explaining how AI decisions may not be fully contestable by human actors.
      \item \textbf{AI Benefits + Assessability Condition}: Includes the AI Benefits Condition plus a paragraph discussing AI’s lack of transparency, including concerns over data privacy and the opacity of its decision-making processes.
\end{itemize}

Note that all AI conditions mention the benefits provided by AI. We pre-registered the expectation that the AI Benefits condition would increase trust in government relative to the human condition. Yet, our theoretical framework posits that overwhelmingly negative reactions emerge once people become more strongly aware of tensions underlying the delegation of public administration tasks to AI, even when people are aware of the benefits of AI use. This yields a non-trivial prediction: although both benefits and drawbacks are present, we expect that negative aspects will dominate in citizens’ evaluations. Compared to the human condition, we expect lower levels of perceived control, trust in government, and AI support in the conditions that highlight dependency, contestability, or assessability risks alongside the benefits. 

We did not pre-register expectations for comparisons among the AI uses (welfare, taxes, bail) and no expectations for the effects of the AI benefits condition on perceived control or AI preferences.

Before data collection, we time‐stamped a detailed pre‐registration plan, specifying theory, hypotheses, research design, power analysis, analysis strategy, analysis code, and falsifiable success criteria (\href{https://osf.io/h4z8g/?view_only=1d1d731a388b463f9148cba1b8fc390c}{OSF preregistration plan}).

\section{Methods and Data}
We recruited participants from the online panel from Prolific, covering a UK sample. The sample is stratified to resemble the UK population in terms of sex, age, and political affiliation. Each participant received 0.9 pounds for participating. The median survey duration was 6.5 minutes.  We collected a sample of 1201 participants as a power simulation indicated that this would give us a power above 0.9 for all of our hypotheses (see \href{https://osf.io/h4z8g/?view_only=1d1d731a388b463f9148cba1b8fc390c}{pre-registration} for the code).\footnote{We aimed for 1,200 participants. The final sample was 1201, as one of the participants finished the survey but was not correctly logged in Prolific's system as a complete case.}

We conducted a pre-test (N=301) to ensure that respondents understood the vignettes as intended. The pre-test used the identical experimental design with added manipulation checks (see \SIref{SI:pretest}). 

For the main study, respondents were recruited via Prolific and redirected to our own survey platform. After providing informed consent, participants were told they would read scenarios describing government operations and subsequently answer related questions. To ensure sufficient engagement with the treatment material, the continue button on each vignette page was disabled for 30 seconds. Following each vignette, participants completed a questionnaire capturing outcome measures. This procedure was repeated three times, with each respondent receiving one vignette on welfare, tax, and bail administration in random order. At the conclusion of the survey, participants were debriefed about the experimental nature of the study and reminded that the vignettes described hypothetical, not necessarily factual, scenarios.

We pre-registered three outcomes. For descriptive statistics on the item level, see \SIref{SI:measures}.
\begin{itemize}
  \item \textbf{Trust in government}, which is a 2-item index (“In this country, you can trust the government to always act in the best interests of its citizens”, “In this country, you can trust the government to make quick and good decisions”). 
  \item \textbf{Perceived loss of control}, which is a 2-item index (“In this country, we are losing control over the important decisions of government“, “In this country, decisions that affect my life are made by anonymous forces beyond my control.").
  \item \textbf{AI Delegation Preference}, which is an ordinal 1-item measure (“Should this state use more or less artificial intelligence, or is the current level just right? More AI, Less AI, Just right”).
\end{itemize}

Our analysis strategy employs multilevel modeling to test the effects of different AI governance conditions on citizens' perceptions of trust in government and feelings of control. We employ linear mixed-effects models to account for the nested structure of our experimental data, where participants evaluate multiple vignettes across different governance domains.\footnote{The code and data to replicate all of our analyses are available on OSF: \href{https://osf.io/ngwrt/files/osfstorage?view_only=b266bf87d89841fc9e6ea83b0855e47f}{Link to data and Code}.} 

Using the lmer function from the lme4 package \cite{bates_fitting_2015}, we fit separate models for the outcome variables trust and sense of control with varying intercepts for the three policy domains (tax administration, welfare benefits, and bail decisions), accounting for individual-level random effects. This approach allows us to estimate the average causal effect of each condition while controlling for baseline variations in trust that might exist across different policy issues and accounts for the non-independence of observations potentially arising from repeated measures within participants. As the outcome variable AI Delegation Preference is ordinal, we use an ordinal regression model also with varying intercepts for the policy domains and the participants and predict mean levels.

As specified in the pre-registration, we excluded from the analysis all respondents who failed both pre-treatment attention checks. This resulted in a final sample of 1,198 participants. To verify the effectiveness of our experimental manipulation, we included a manipulation check after each vignette, asking participants to identify the policy domain described. The results indicate exceptionally high recall rates (tax: 99.1\%, welfare: 99.0\%, bail: 98.8\%), suggesting that respondents engaged closely with the material.

\section{Results}
\begin{figure}[!htb]
\centering
\includegraphics[width=\textwidth,height=\textheight,keepaspectratio]{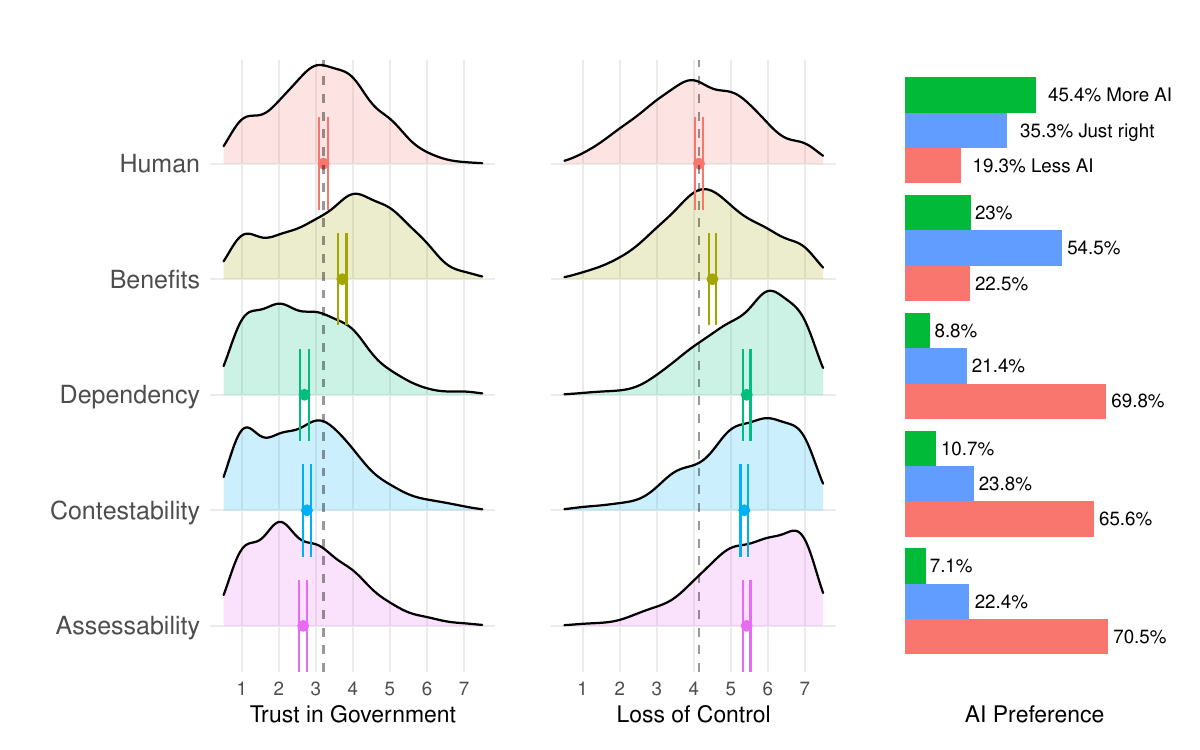}
\caption{Density and bar plots for all three dependent variables with the treatment effects. The dashed line in the first two panels indicates the mean score of the human condition. For each condition, the estimates with 95\%-CIs are shown.}
\label{fig:fig_results}
\end{figure}

Figure \ref{fig:fig_results} presents the aggregated results across all experimental conditions, outcomes, and hypothesis tests. The dashed vertical line marks the mean level in the human condition, which we pre-registered as the baseline for all comparisons. 

First, we assess participants' responses to scenarios in which governmental tasks are either fully managed by human officials or assisted by AI systems. AI systems were described as handling routine tasks and providing tangible benefits. We expected that framing AI deployment for routine tasks as providing tangible benefits would enhance trust in government compared to purely human management (see left panel of Figure \ref{fig:fig_results}). We had no definitive expectations regarding whether such narrow AI use would affect participants' perceived sense of control (middle panel of Figure \ref{fig:fig_results}) or their demand for broader AI adoption in government (right panel of Figure \ref{fig:fig_results}).

The comparison between the human and AI benefits conditions aligns with our predictions: awareness of the tangible advantages provided by AI in routine tasks does indeed enhance trust in government. This effect is consistent across all three domains (for the complete tables of the models see \SIref{SI:models}), yielding an average increase of 0.5 scale points (p<0.001; 95\% CI [0.40, 0.62]), indicative of a significant yet moderate increase in trust, from 3.21 to 3.71. Thus, while AI positively influences government trust, it does not radically alter trust levels ($\beta$=0.36; 95\% CI [0.28, 0.43]), which remain moderate overall.

The right panel shows that 45.4\% of respondents in the human condition prefer delegating more governmental tasks to AI, highlighting an openness to technological integration. The scenario highlighting the narrow, beneficial use of AI largely seems to meet this demand, reducing the proportion of respondents advocating for even broader AI use to just 23\%.

Nevertheless, even the beneficial framing of AI use is not perceived as unequivocally positive. We observe a corrosive impact of AI on participants' feelings of control (b=0.36; p<0.001; 95\% CI [0.24, 0.48]; $\beta$=0.25). This finding is robust across all tested domains (see appendix), suggesting that even optimal AI implementations inherently erode perceived control. This result can be interpreted as evidence supporting the uncanny valley hypothesis, which posits that the impersonal nature of AI in any implementation may be perceived as alien and detrimental to the feeling of being in control.

Consequently, even narrow AI implementations delivering tangible benefits may not represent Pareto-optimal solutions but instead pose substantial trade-offs with respect to public opinion. On the one hand, citizens expect governments to modernize and utilize AI to enhance efficiency in routine tasks, and indeed, such implementations bolster governmental trust. On the other hand, these benefits are inseparable from a pronounced loss of perceived control. Given that trust and perceived control are both foundational to democratic legitimacy, the integration of AI into governmental tasks emerges as inherently ambivalent, simultaneously strengthening and weakening critical pillars of democracy.

The remaining experimental conditions test our primary hypotheses derived from principal-agent theory, which suggested that awareness of the tensions inherent in AI delegation undermines perceived system legitimacy. To evaluate these implications, we compare the dependency, contestability, and assessability conditions against the human condition.

We find that all three AI conditions significantly erode institutional trust relative to the human condition (Figure \ref{fig:fig_results}). In other words, awareness of each of the tensions identified through PA theory has the expected corrosive effects on both institutional trust and perceived loss of control. The magnitude of this erosion is similar across all conditions. 

The effects are particularly strong for people’s sense of losing control (dependency $\beta$=0.89; 95\% CI [0.80, 0.97]; contestability $\beta$=0.84; 95\% CI [0.76, 0.92]; assessability $\beta$=0.88; 95\% CI [0.80, 0.97]). Specifically, on a 1-7 scale, the perceived loss of control rises from an average of 4.14 in the human condition to 5.43 in the dependency condition, 5.36 in the contestability condition, and 5.43 in the assessability condition. Dichotomizing the outcome variables at their scale mid-point provides another angle on the strength of the observed effects. While 45.4\% of respondents in the human condition report feeling a lack of control (above the scale midpoint), this proportion sharply increases to 81.6\% within the dependency condition.

Unsurprisingly, highlighting the risks inherent to task delegation also has substantial effects on people's preferences towards delegating tasks to AI. Across the three dimensions of risk,  between 65.6\% and 70.5\% of respondents demand less AI when these risks are made salient. Collectively, these findings substantiate the expectations derived from principal-agent theory that awareness of predictable tensions in delegating tasks to AI yields corrosive outcomes. Overall, our data supports all three hypotheses.

To substantiate the processes underlying these results, we analyse data from our pre-test, which we had fielded before the main study among 301 respondents from the same Prolific panel. The pre-test included three manipulation checks to test whether the observed effects occur through the theoretically anticipated attitudinal mechanisms (for the full models, see \SIref{SI:pretest}). 

For the dependency condition, we expected respondents to perceive an irreversible dependency on AI, arising from an erosion of human capacity in government. Supporting this, respondents in the dependency condition in contrast to the status quo condition exhibited significantly stronger agreement with the statement: "The government in this country could no longer function effectively without relying on AI systems" (b=1.57; p<0.001; 95\% CI [1.28, 1.87]; $\beta$=0.89), which precedes reductions in institutional trust and increases in perceived loss of control.

Respondents in the assessability condition were made aware of AI’s limited transparency, highlighting concerns regarding data privacy and decision-making opacity. Consistent with expectations, these participants displayed stronger agreement with the statement: "In this country, it is difficult to scrutinize how the government reaches its decisions" (b=1.00; p<0.001; 95\% CI [0.73, 1.28]; $\beta$=0.62), which predicts subsequent erosion in institutional trust and control perceptions.

Finally, participants in the contestability condition faced a scenario emphasizing the difficulty of holding AI accountable, particularly when decisions are erroneous. Aligning with our predictions, respondents expressed lower agreement with the statement: "If I feel that I have been treated unfairly by the state, I have the opportunity to effectively challenge a state decision" (b=-0.90; p<0.001; 95\% CI [-1.10, -0.51]; $\beta$=-0.55), reinforcing observed declines in trust and perceptions of control.

\section{Conclusion}
Our findings show that the use of AI in public administration tasks has distinct and measurable effects on public perceptions. When participants are made aware of AI's potential to improve efficiency in routine governmental tasks, institutional trust increases moderately. However, even under these favorable conditions, participants report a reduced sense of control. In contrast, when participants are made aware of the structural challenges associated with delegating tasks to AI\textemdash specifically dependency, contestability, and assessability\textemdash both perceived control and institutional trust decline substantially. These patterns suggest that while the prospect of efficiency can enhance trust, awareness of underlying governance challenges systematically reduces public confidence and the feeling of democratic control.

These findings align with findings from other uses of AI in politics, where we clearly also find evidence of penalties for AI uses, be it for deliberation and participation or political communication \cite{jungherr_deceptive_2024, jungherr_artificial_2025}. People tend to react negatively toward AI use in politics. This points to several underlying challenges in the governmental use of AI. Even when AI is perceived as efficient and beneficial in improving public service delivery, people report a diminished sense of control when informed about its use. This perceived control loss is especially concerning in democratic contexts, where trust and agency are foundational. In times when some liberal democracies face declining public support and legitimacy \cite{valgarsson_crisis_2025}, prioritizing efficiency at the expense of perceived democratic control may risk reinforcing these broader trends. Careful attention is therefore needed to ensure that technological innovation does not inadvertently deepen democratic disaffection.

A further, more forward-looking concern emerges from our findings. While initial uses of AI in government may be accepted by the public, broader implementation is likely to increase public awareness of the technology’s inner workings and associated governance challenges. Although we could not survey future respondents directly, our information treatments approximated such a scenario by exposing participants to expected issues related to AI delegation. The results indicate that this broader awareness not only reduces individuals’ sense of control but also diminishes institutional trust and decreases public support for expanding AI use in public administration. This pattern suggests a potential risk of "failure through success," where early efficiency gains give way to legitimacy losses as AI use widens and public understanding deepens.

The paper also demonstrates the utility of principal-agent theory for identifying tensions in the delegation of public sector tasks to AI. Principal-agent theory highlights two core problems of delegation that are especially salient when tasks are transferred to AI systems: information asymmetry \cite{holmstrom_moral_1979, miller_political_2005} and power reversals \cite{abbott_competencecontrol_2020}. These mechanisms reduce the ability of administrative supervisors and of citizens, as principals,  to monitor or influence decision-making by AI systems, thereby undermining perceptions of control. Drawing on this framework, we developed testable hypotheses and designed experimental treatments that captured anticipated future concerns about dependency, contestability, and assessability. Our findings support the validity of this approach: participants responded to these scenarios in ways that align with the theory’s predictions. In this context, principal-agent theory functioned not just as an explanatory framework but as a predictive tool for anticipating likely sources of public resistance to AI-driven administrative delegation.

Importantly, our principal–agent framework for AI can be extended to other settings facing threats similar to those outlined in our study. Future research could apply this framework across organizational contexts where AI systems are deployed to examine how dependency, contestability, and assessability shape domain-specific perceptions.

Of course, our study has limitations. It relies on hypothetical vignette scenarios rather than real-world implementations. As such, external validity may be constrained, and actual political responses could differ when AI is embedded within visible public institutions. Additionally, although our participant sample was stratified to reflect the UK population on key demographic and political variables, it was not a probability sample. These factors may limit the generalizability of our results.

Future research should explore how reactions to governmental AI evolve over time as the technology becomes more familiar and integrated into daily life. Comparative studies across political cultures could shed light on how democratic norms, institutional trust, and attitudes toward technology interact with information of governmental uses of AI. Furthermore, studies examining citizens' direct experience with actual AI-enabled systems in government contexts could provide valuable insights beyond hypothetical scenarios.

In short, while AI offers real opportunities for more efficient public administration, our findings show that perceived legitimacy and democratic control cannot be taken for granted. Anticipating and addressing public concerns, particularly those tied to control, accountability, and transparency, will be essential to ensuring that the future of digital governance does not lead to a greater sense of distance between people and their democratic government.

\section*{Acknowledgments}
\label{acknowledgements}
Alexander Wuttke gratefully acknowledges funding from the LMU Young Researcher Fund. Adrian Rauchfleisch’s work was supported by the  National Science and Technology Council, Taiwan (R.O.C) (Grant No 113-2628-H-002-018-).

\bibliographystyle{apacite}
\bibliography{principal_agent}

\clearpage

\appendix
\appendix           
\part{}             
\parttoc            

\section{Deviations from pre-analysis plan}
\label{SI:deviations}
\begin{itemize}
  \item Compared to the original plans, our sample has one additional respondent as Prolific missed logging one completed interview.
  \item Compared to the pre-registered analysis script, we modified the visual presentation of the effect estimates while the underlying models remain unchanged.
\end{itemize}

\section{Measures}
\label{SI:measures}
A copy of the complete questionnaire is available on OSF (\href{https://osf.io/ngwrt/files/osfstorage?view_only=b266bf87d89841fc9e6ea83b0855e47f}{questionnaire}).
In this section, we show the vignettes used as treatments in our experiment. In the second sub-section, we report the complete descriptive statistics on the item level for all variables.

\subsection{Vignettes used as treatments}
\label{SI:vignettes}
\subsubsection{AI in Tax Administration}
\begin{description}[font=\bfseries,leftmargin=0pt,labelsep=1em]
  \item[Human Condition]
Imagine that your country's tax administration works as follows:
Tax returns are processed primarily by trained tax officials. For digitally submitted tax returns, computer algorithms provide support for simple verification tasks, but human intervention is required in particular for paper tax returns. Each tax return is then checked by hand. Routine tasks in particular take a lot of time. It takes a long time to check by hand whether all mandatory fields have been filled out or whether obviously nonsensical values have been entered, such as decimal places that have been shifted due to careless mistakes. Because these routine tasks are tedious, people sometimes make mistakes. On the other hand, it is time-consuming. That is why it sometimes takes a long time for tax refunds to be paid out.

  \item[AI Benefits Condition]
Imagine that your country's tax administration works as follows:
Each tax return is first checked by an AI system. The AI can analyze the entire tax return in seconds and automatically determine whether all mandatory fields have been completed and whether the entries are plausible. It immediately detects typos, such as misplaced decimal points or unusual values. AI support significantly reduces processing time. Tax refunds are issued much faster than with purely human processing. In addition, the use of AI reduces administrative costs, which contributes to more efficient public finances in the long term.

  \item[AI Benefits Condition + Dependency Condition]
Imagine that your country's tax administration works as follows:
Each tax return is first checked by an AI system. The AI can analyse the entire tax return in seconds and automatically determine whether all mandatory fields have been completed and whether the entries are plausible. It immediately detects typos such as misplaced decimal points or unusual values. AI support significantly reduces processing time. Tax refunds are issued much faster than with purely human processing. In addition, the use of AI reduces administrative costs, which contributes to more efficient public finances in the long term.
With the introduction of AI, numerous human tax auditors were laid off, which was initially welcomed as a sensible cost saving. However, problems have since emerged: the remaining specialists can hardly check or correct the decisions of the AI. When systematic errors recently occurred in the processing of tax returns from self-employed individuals, the necessary expertise to identify the cause was lacking. Experts are now warning that the state has become dangerously dependent on AI. Human expertise has been so greatly reduced that a return to human auditing procedures would hardly be possible – both the sufficiently qualified specialists and the knowledge gained through years of experience are lacking. Without sufficient human expertise, the tax authority is at the mercy of the AI system – even if it turns out that it makes mistakes or can be manipulated. The authority is now virtually forced to trust the AI, even if its decisions appear flawed or non-transparent.

  \item[AI Benefits Condition + Contestability Condition]
Imagine that your country's tax administration works as follows:
Each tax return is first checked by an AI system. The AI can analyse the entire tax return in seconds and automatically determine whether all mandatory fields have been completed and whether the entries are plausible. It immediately detects typos such as misplaced decimal points or unusual values. AI support significantly reduces processing time. Tax refunds are issued much faster than with purely human processing. In addition, the use of AI reduces administrative costs, which contributes to more efficient public finances in the long term.
Although officially all tax assessments must be signed by a human auditor, the reality has changed fundamentally. The most important decisions, those that have a significant financial impact on citizens, are actually made by an anonymous machine. After extensive staff cuts, the remaining tax officials are confronted with such a flood of AI-generated assessments that careful human review has become impossible. Most AI decisions are therefore adopted without review – the human becomes a mere rubber stamp for what the machine proposes. If a tax assessment is incorrect or someone is treated unfairly, there is a lack of human auditors to process objections. Because there is a shortage of tax auditors and people generally trust the AI, the first review of objections is also carried out by the AI. It is only when a citizen files a lawsuit in the second instance that there is a right to review by a human tax clerk. Since this process is expensive and lengthy for citizens, only a few AI decisions are corrected by human review, even if they are incorrect . What was intended as progress turns into a dehumanised bureaucratic apparatus.

  \item[AI Benefits Condition + Assessability Condition]
Imagine that your country's tax administration works as follows:
Each tax return is first checked by an AI system. The AI can analyse the entire tax return in seconds and automatically determine whether all mandatory fields have been completed and whether the entries are plausible. It immediately detects typos such as misplaced decimal points or unusual values. AI support significantly reduces processing time. Tax refunds are issued much faster than with purely human processing. In addition, the use of AI reduces administrative costs, which contributes to more efficient public finances in the long term.
But no one knows exactly which criteria the AI actually uses to make decisions. No one can understand which criteria the AI actually uses to make decisions. The system's manufacturer stubbornly refuses to disclose which data and which rules the AI was trained with – allegedly to protect trade secrets. Even the tax administration itself is not entirely sure. Critics warn that the AI could favour wealthy taxpayers by evaluating their tax returns more generously, while others are audited more rigorously. This has led to accusations that certain population groups are being systematically favoured or disadvantaged. But because no one can look into the AI's decision-making logic, suspicions remain – independent verification is simply not possible. Nor are the AI's individual decisions comprehensible. The machine merely states that a tax return is correct or incorrect – but it gives no reasons, no indications of the considerations it has made.

\end{description}

\subsubsection{AI in Welfare Benefits Approval}
\begin{description}[font=\bfseries,leftmargin=0pt,labelsep=1em]
  \item[Human Condition]
Imagine the welfare benefits system in your country works as follows:
Applications for welfare benefits are primarily processed by trained social welfare officials. Each application requires extensive documentation, including income statements, residence verification, and family status information. Officials manually review each document, cross-reference information with existing databases, and assess eligibility against complex criteria. This manual processing takes considerable time, often 4-6 weeks, for a standard application. The complexity of regulations and the volume of applications sometimes lead to inconsistencies in decision-making between different officials. Human error occasionally results in overlooked documentation or misinterpreted eligibility criteria. Applicants frequently experience delays in benefit payments, which can cause significant hardship for those in urgent financial need. Additionally, the system struggles during periods of increased applications, such as economic downturns, when processing times can extend to several months.

\item[AI Benefits Condition]
Imagine the welfare benefits system in your country works as follows:
Every welfare application is processed through an AI system. The AI can analyze all submitted documents within minutes, automatically verifying information against multiple government databases. The system instantly identifies missing documentation and notifies applicants about what additional information is needed. It applies eligibility criteria consistently across all applications, eliminating human bias and regional variations in decision-making. Through the AI-supported process, application assessment time has been reduced from weeks to just days, with straightforward cases often approved within 24 hours. Emergency assistance can be prioritized and processed even faster. The AI system also detects potential fraud patterns more effectively than human reviewers, reducing improper payments. Additionally, the implementation of AI has significantly reduced administrative costs, allowing more resources to be directed toward benefit payments and support services for recipients.

\item[AI Benefits + Dependency Condition]
Imagine the welfare benefits system in your country works as follows:
Every welfare application is processed through an AI system. The AI can analyze all submitted documents within minutes, automatically verifying information against multiple government databases. The system instantly identifies missing documentation and notifies applicants about what additional information is needed. It applies eligibility criteria consistently across all applications, eliminating human bias and regional variations in decision-making. Through the AI-supported process, application assessment time has been reduced from weeks to just days, with straightforward cases often approved within 24 hours. Emergency assistance can be prioritized and processed even faster. The AI system also detects potential fraud patterns more effectively than human reviewers, reducing improper payments. Additionally, the implementation of AI has significantly reduced administrative costs, allowing more resources to be directed toward benefit payments and support services for recipients.
With the introduction of AI, numerous social welfare officials were laid off, initially welcomed as a cost-saving measure. However, problematic consequences have emerged: The remaining staff can barely review or correct the AI's decisions. When systematic errors occurred in processing applications from self-employed individuals with irregular income patterns, the necessary expertise to identify the cause was lacking. Experts now warn that the state has entered a dangerous dependency on the AI system. Human expertise has been so significantly reduced that returning to human assessment procedures would be nearly impossible – there are neither enough qualified professionals nor the experiential knowledge lost over the years. Without sufficient human expertise, the welfare agency remains at the mercy of the AI system – even if it makes errors or could be manipulated. The agency is now essentially forced to trust the AI, even when its decisions appear flawed or non-transparent.

\item[AI Benefits + Contestability Condition]
Imagine the welfare benefits system in your country works as follows:
Every welfare application is processed through an AI system. The AI can analyze all submitted documents within minutes, automatically verifying information against multiple government databases. The system instantly identifies missing documentation and notifies applicants about what additional information is needed. It applies eligibility criteria consistently across all applications, eliminating human bias and regional variations in decision-making. Through the AI-supported process, application assessment time has been reduced from weeks to just days, with straightforward cases often approved within 24 hours. Emergency assistance can be prioritized and processed even faster. The AI system also detects potential fraud patterns more effectively than human reviewers, reducing improper payments. Additionally, the implementation of AI has significantly reduced administrative costs, allowing more resources to be directed toward benefit payments and support services for recipients.
Although, officially, all benefit decisions require approval by a human supervisor, the reality has significantly changed. Critical decisions about essential financial support for vulnerable citizens are now effectively determined by an impersonal AI system. Due to substantial workforce reductions, the remaining welfare officers are overwhelmed by the volume of AI-generated decisions, making thorough human oversight nearly impossible. Most AI decisions are approved without meaningful review. If the AI incorrectly denies benefits or calculates incorrect amounts, citizens encounter substantial barriers to challenging these decisions. Initially, applicants must navigate a complicated and lengthy AI-based appeals process. Human reconsideration is only available after an initial AI review, due to staff shortages and strong institutional reliance on the AI system. Applicants seeking further review must pursue a second-level formal complaint—a complex process demanding time, resources, and expertise typically unavailable to disadvantaged individuals. This creates significant frustration and hardship, resulting in infrequent correction of erroneous AI decisions. Consequently, a system initially intended as progress has evolved into a bureaucratic barrier, limiting meaningful recourse for those it aimed to support.

\item[AI Benefits + Assessability Condition]
Imagine the welfare benefits system in your country works as follows:
Every welfare application is processed through an AI system. The AI can analyze all submitted documents within minutes, automatically verifying information against multiple government databases. The system instantly identifies missing documentation and notifies applicants about what additional information is needed. It applies eligibility criteria consistently across all applications, eliminating human bias and regional variations in decision-making. Through the AI-supported process, application assessment time has been reduced from weeks to just days, with straightforward cases often approved within 24 hours. Emergency assistance can be prioritized and processed even faster. The AI system also detects potential fraud patterns more effectively than human reviewers, reducing improper payments. Additionally, the implementation of AI has significantly reduced administrative costs, allowing more resources to be directed toward benefit payments and support services for recipients.
Yet nobody knows exactly what criteria the AI uses to make its decisions. No one can trace how the AI actually evaluates applications or weighs different factors. The system developer persistently refuses to disclose what data and rules were used to train the AI – allegedly to protect trade secrets. Even the welfare administration itself doesn't know precisely how decisions are made. Critics warn that the AI might favor certain demographic groups while scrutinizing others more harshly. There are accusations that single parents, minorities, or applicants from certain neighborhoods might be systematically disadvantaged, while others receive preferential treatment. But because no one can examine the AI's decision-making logic, these remain suspicions – independent verification is simply impossible. Individual AI decisions are equally opaque. The system merely communicates that an application is approved or denied – but provides no reasoning or explanation of what considerations influenced the outcome. Applicants receive standardized rejection letters with generic reasons, leaving them unable to understand specifically why their application was denied or what they could change to become eligible.

\end{description}

\subsubsection{AI in Risk Assessment for Bail Decisions}
\begin{description}[font=\bfseries,leftmargin=0pt,labelsep=1em]
  \item[Human Condition]
Imagine the bail decision process in your country works as follows: 
Bail decisions are made by judges who review each defendant’s case manually. Judges consider various factors such as criminal history, severity of charges, employment status, and community ties. Due to the complexity of cases and individual differences among judges, decisions on who receives bail and under what conditions can vary widely. Cognitive biases, fatigue, and varying interpretations of criteria often lead to inconsistent and unpredictable outcomes. This inconsistency can result in unequal treatment of defendants facing similar charges and circumstances, leading to public perceptions of unfairness and reduced trust in the judicial system.

\item[AI Benefits Condition]
Imagine the bail decision process in your country works as follows: 
Every defendant’s bail eligibility is initially assessed by an AI-driven system that generates a risk score based on extensive data analysis. This AI system consistently evaluates standardized criteria such as criminal record, charges, and personal circumstances using historical and demographic data. The systematic nature of AI-driven risk scoring significantly reduces human biases and ensures greater consistency across different cases. As a result, bail decisions have become more predictable and uniform, enhancing fairness and transparency. Additionally, this data-driven approach allows courts to process cases more efficiently, speeding up hearings and reducing pre-trial detention periods for defendants.

\item[AI Benefits + Dependency Condition]
Imagine the bail decision process in your country works as follows: 
Every defendant’s bail eligibility is initially assessed by an AI-driven system that generates a risk score based on extensive data analysis. This AI system consistently evaluates standardized criteria such as criminal record, charges, and personal circumstances using historical and demographic data. The systematic nature of AI-driven risk scoring significantly reduces human biases and ensures greater consistency across different cases. As a result, bail decisions have become more predictable and uniform, enhancing fairness and transparency. Additionally, this data-driven approach allows courts to process cases more efficiently, speeding up hearings and reducing pre-trial detention periods for defendants.
However, the widespread use of AI for bail decisions has caused judges to rely excessively on AI-generated risk scores. Judges increasingly defer to the AI’s recommendations without independently verifying or questioning the decisions.  What began as a supportive tool has evolved into the de facto decision-maker, with judges increasingly deferring to the algorithmic recommendations without substantive scrutiny. New judges receive minimal training on independently evaluating flight risk and public safety, instead learning primarily how to interpret AI outputs. Over time, this dependency has eroded judges' ability to critically assess and challenge AI evaluations, particularly in nuanced or borderline cases. Experts warn that the judicial system risks losing critical human judgment and legal expertise, becoming overly dependent on automated decisions that may occasionally be flawed or biased.

  \item[AI Benefits + Contestability Condition]
Imagine the bail decision process in your country works as follows: 
Every defendant’s bail eligibility is initially assessed by an AI-driven system that generates a risk score based on extensive data analysis. This AI system consistently evaluates standardized criteria such as criminal record, charges, and personal circumstances using historical and demographic data. The systematic nature of AI-driven risk scoring significantly reduces human biases and ensures greater consistency across different cases. As a result, bail decisions have become more predictable and uniform, enhancing fairness and transparency. Additionally, this data-driven approach allows courts to process cases more efficiently, speeding up hearings and reducing pre-trial detention periods for defendants.
Yet, AI-generated bail decisions are extremely difficult for defendants to challenge effectively. If a defendant disagrees with the AI-generated risk assessment, they must undergo a complicated appeals process. The courts presume the AI system's technical correctness, placing the burden on defendants to prove why the algorithmic assessment is flawed in their specific case. Due to reduced judicial oversight and institutional trust in the AI system, human review is rarely available. Defendants wishing to challenge their bail conditions or denial face lengthy and resource-intensive legal procedures.  Appeals based on challenging the AI assessment face strict procedural hurdles and extended timelines, rendering them impractical for pretrial detention decisions that require immediate resolution. As a result, incorrect or unjustified AI assessments are seldom overturned, causing significant frustration and potential injustice for affected defendants.

  \item[AI Benefits + Assessability Condition]
Imagine the bail decision process in your country works as follows: 
Every defendant’s bail eligibility is initially assessed by an AI-driven system that generates a risk score based on extensive data analysis. This AI system consistently evaluates standardized criteria such as criminal record, charges, and personal circumstances using historical and demographic data. The systematic nature of AI-driven risk scoring significantly reduces human biases and ensures greater consistency across different cases. As a result, bail decisions have become more predictable and uniform, enhancing fairness and transparency. Additionally, this data-driven approach allows courts to process cases more efficiently, speeding up hearings and reducing pre-trial detention periods for defendants.
However, the exact criteria and decision-making processes used by the AI system remain opaque to the public, defendants, and even court officials. The developer of the AI system refuses to disclose specific training data and decision-making algorithms, citing proprietary rights and security concerns. Due to this lack of transparency, concerns have arisen regarding potential systematic biases or discrimination embedded in the AI's evaluations. Critics warn that certain demographic groups might be unfairly assessed as higher risks, leading to disproportionate pre-trial detention rates. Without clear insights into the AI’s logic, these concerns cannot be independently verified, undermining public trust and defendants’ confidence in the fairness of the judicial system.

\end{description}

\subsection{Complete descriptive table with all variables and items}
\label{section:items}


\begin{table}[H]
\resizebox{\textwidth}{!}{
\begin{tabular}{L{4cm}L{5cm}cc}
\toprule
Variable & Question/Operationalization & M (SD) & n\\
\midrule
Trust in government (2 items, $\alpha$ = 0.85-0.88, Spearman-Brown = 0.85-0.88) & (1="Strongly disagree", 7="Strongly agree") & 3.01 (1.43) & 3594\\
 & In this country, you can trust the government to always act in the best interests of its citizens. & 3.01 (1.53) & 3594\\
 &  In this country, you can trust the government to make quick and good decisions. & 3.02 (1.52) & 3594\\

Control loss (2 items, $\alpha$ = 0.74-0.76, Spearman-Brown =0.74-0.76) & (1="Strongly disagree", 7="Strongly agree") & 4.96 (1.45) & 3594\\
 & In this country, we are losing control over the important decisions of government. & 4.86 (1.65) & 3594\\
 & In this country, decisions that affect my life are made by anonymous forces beyond my control. & 5.07 (1.62) & 3594\\

AI preference & Should this government use more or less artificial intelligence, or is the current level just right? & & 3594\\
 & Less AI & 49.0\% & \\
 & Just right & 31.7\% & \\
 & More AI & 19.3\% & \\

Treatments conditions & &  & 3594\\
 & Human & & 736\\
 & AI Positive &  & 743\\
 & AI dependency &  & 706\\
 & AI contestability &  & 694\\
 & AI Assessability  &  & 715\\

Treatments issues & &  & 3594\\
 & Tax & & 1198\\
 & Welfare &  & 1198\\
 & Bail &  & 1198\\

Age & (in years) & 46.76 (15.7) & 1198\\
Gender & (1 = male) & 48.3\% & 1198\\
Education & (1 = Postgraduate degree) & 19.3\% & 1198\\
 
\bottomrule
\end{tabular}
}
\caption{Descriptive statistics for all relevant variables and items. We calculated the reliability scores for each issue.}
\label{tab:descr_01}
\end{table}

\section{Model results}
\label{SI:models}
In this section, we report all the models from the main paper.

\begin{table}[H]
\begin{tabular}[t]{lcccc}
\toprule
Predictors & Estimate & LL & UL & p\\
\midrule
Intercept             & 3.21  & 3.09 & 3.32  & <0.001\\
AI Positive           & 0.51  & 0.40 & 0.62  & <0.001\\
AI Dependency         & -0.51 & -0.63 & -0.40 & <0.001\\
AI Contestability     & -0.45 & -0.56 & -0.34 & <0.001\\
AI Assessability      & -0.55 & -0.66 & -0.44 & <0.001\\
\midrule
\multicolumn{5}{l}{\textbf{Random Effects}}\\
$\sigma^2$                 & 0.85  &      &      &       \\
$\tau_{00}$ participant    & 1.04  &      &      &       \\
$\tau_{00}$ issue          & 0.00  &      &      &       \\
ICC                        & 0.55  &      &      &       \\
\midrule
N issue                    & 3     &      &      &       \\
N participant              & 1198  &      &      &       \\
Observations               & 3594  &      &      &       \\
Marginal $R^2$ / Conditional $R^2$ & 0.081/0.587 &      &      &       \\
\bottomrule
\end{tabular}
\caption{Mixed‐effects model for the dependent variable trust in government with 95\% confidence intervals.}
\end{table}

\begin{table}[H]
\begin{tabular}[t]{lcccc}
\toprule
Predictors & Estimate & LL & UL & p\\
\midrule
Intercept           & 4.14  & 4.04 & 4.24 & <0.001\\
AI Positive         & 0.36  & 0.24 & 0.48 & <0.001\\
AI Dependency       & 1.29  & 1.17 & 1.41 & <0.001\\
AI Contestability   & 1.22  & 1.10 & 1.34 & <0.001\\
AI Assessability    & 1.29  & 1.17 & 1.41 & <0.001\\
\midrule
\multicolumn{5}{l}{\textbf{Random Effects}}\\
$\sigma^2$               & 1.03  &      &      &       \\
$\tau_{00}$ participant     & 0.77  &      &      &       \\
$\tau_{00}$ issue        & 0.00  &      &      &       \\
ICC                      & 0.43  &      &      &       \\
\midrule
N issue                  & 3     &      &      &       \\
N participant              & 1198  &      &      &       \\
Observations             & 3594  &      &      &       \\
Marginal $R^2$ / Conditional $R^2$ & 0.142/0.509 &      &      &       \\
\bottomrule
\end{tabular}
\caption{Mixed‐effects model for the dependent variable control loss with 95\% confidence intervals.}
\end{table}

\begin{table}[H]
\begin{tabular}[t]{lcccc}
\toprule
Predictors & Odds Ratios & LL & UL & p\\
\midrule
1-less AI|2-just right     & 0.09  & 0.07 & 0.12 & <0.001\\
2-just right|3-More AI     & 1.58  & 1.20 & 2.07 & 0.001\\
AI Positive                  & 0.36  & 0.27 & 0.46 & <0.001\\
AI Dependency                & 0.03  & 0.02 & 0.04 & <0.001\\
AI Contestability            & 0.03  & 0.02 & 0.05 & <0.001\\
AI Assessability             & 0.03  & 0.02 & 0.04 & <0.001\\
\midrule
\multicolumn{5}{l}{\textbf{Random Effects}}\\
$\sigma^2$                   & 3.29  &      &      &       \\
$\tau_{00}$ participant         & 4.02  &      &      &       \\
$\tau_{00}$ issue            & 0.02  &      &      &       \\
ICC                          & 0.55  &      &      &       \\
\midrule
N issue                      & 3     &      &      &       \\
N participant                   & 1198  &      &      &       \\
Observations                 & 3594  &      &      &       \\
Marginal $R^2$ / Conditional $R^2$ & 0.246/0.661 &      &      &       \\
\bottomrule
\end{tabular}
\caption{Cumulative Link mixed model results for the ordinal dependent variable AI preference: thresholds 1-less AI|2-just right, 2-just right|3-More AI), odds ratios with lower (LL) and upper (UL) limits, and random‐effect variance components.}
\end{table}

\section{Single case analysis}
In this section, we report single models for each case as a robustness analysis.

\subsection{Tax case}
\begin{table}[H]
\centering
\begin{tabular}[t]{lcccc}
\toprule
Predictors                  & Estimate & LL    & UL    & p       \\
\midrule
Intercept                   & 3.31     & 3.14  & 3.48  & <0.001  \\
AI Positive          & 0.41     & 0.17  & 0.65  & 0.001   \\
AI Dependency         & -0.55    & -0.79 & -0.31 & <0.001  \\
AI Contestability    & -0.57    & -0.82 & -0.32 & <0.001  \\
AI Assessability     & -0.65    & -0.90 & -0.41 & <0.001  \\
\midrule
Observations                & 1198     &       &       &         \\
R$^2$ / R$^2$ adjusted      & 0.084 / 0.080 & &    &         \\
\bottomrule
\end{tabular}
\caption{Dependent variable is trust in government. Ordinary least squares regression model with 95\% confidence intervals.}
\end{table}

\begin{table}[H]
\centering
\begin{tabular}[t]{lcccc}
\toprule
Predictors                  & Estimate & LL    & UL    & p       \\
\midrule
Intercept                   & 3.98     & 3.81  & 4.14  & <0.001  \\
AI Positive          & 0.35     & 0.11  & 0.59  & 0.004   \\
AI Dependency         & 1.50     & 1.26  & 1.73  & <0.001  \\
AI Contestability    & 1.44     & 1.20  & 1.68  & <0.001  \\
AI Assessability     & 1.50     & 1.26  & 1.73  & <0.001  \\
\midrule
Observations                & 1198     &       &       &         \\
R$^2$ / R$^2$ adjusted      & 0.193 / 0.191 &   &    &         \\
\bottomrule
\end{tabular}
\caption{Dependent variable is loss of control. Ordinary least squares regression model with 95\% confidence intervals.}
\end{table}

\begin{table}[H]
\centering
\begin{tabular}[t]{lcccc}
\toprule
Predictors                  & Odds Ratios & LL    & UL    & p       \\
\midrule
Less AI $|$ Just right      & 0.17        & 0.13  & 0.23  & <0.001  \\
Just right $|$ More AI      & 1.22        & 0.95  & 1.56  & 0.113   \\
AI Positive          & 0.58        & 0.42  & 0.81  & 0.001   \\
AI Dependency         & 0.08        & 0.06  & 0.12  & <0.001  \\
AI Contestability    & 0.09        & 0.06  & 0.14  & <0.001  \\
AI Assessability     & 0.07        & 0.05  & 0.11  & <0.001  \\
\midrule
Observations                & 1198        &       &       &         \\
R$^2$ Nagelkerke            & 0.290       &       &       &         \\
\bottomrule
\end{tabular}
\caption{Ordinal logistic regression model with odds ratios and 95\% confidence intervals.}
\end{table}

\subsection{Welfare case}
\begin{table}[H]
\centering
\begin{tabular}[t]{lcccc}
\toprule
Predictors                  & Estimate & LL    & UL    & p       \\
\midrule
Intercept                   & 2.97     & 2.80  & 3.14  & <0.001  \\
AI Positive          & 0.67     & 0.43  & 0.91  & <0.001  \\
AI Dependency         & -0.24    & -0.49 & 0.01  & 0.056   \\
AI Contestability    & -0.20    & -0.44 & 0.05  & 0.114   \\
AI Assessability     & -0.40    & -0.66 & -0.15 & 0.002   \\
\midrule
Observations                & 1198     &       &       &         \\
R$^2$ / R$^2$ adjusted      & 0.071 / 0.068 & &   &         \\
\bottomrule
\end{tabular}
\caption{Dependent variable is trust in government. Ordinary least squares regression model with 95\% confidence intervals.}
\end{table}

\begin{table}[H]
\centering
\begin{tabular}[t]{lcccc}
\toprule
Predictors                  & Estimate & LL    & UL    & p       \\
\midrule
Intercept                   & 4.24     & 4.07  & 4.41  & <0.001  \\
AI Positive          & 0.23     & -0.01 & 0.46  & 0.056   \\
AI Dependency         & 1.20     & 0.96  & 1.44  & <0.001  \\
AI Contestability    & 1.14     & 0.91  & 1.38  & <0.001  \\
AI Assessability     & 1.34     & 1.09  & 1.58  & <0.001  \\
\midrule
Observations                & 1198     &       &       &         \\
R$^2$ / R$^2$ adjusted      & 0.148 / 0.145 &   &    &         \\
\bottomrule
\end{tabular}
\caption{Dependent variable is loss of control. Ordinary least squares regression model with 95\% confidence intervals.}
\end{table}

\begin{table}[H]
\centering
\begin{tabular}[t]{lcccc}
\toprule
Predictors                  & Odds Ratios & LL    & UL    & p       \\
\midrule
Less AI $|$ Just right      & 0.14        & 0.10  & 0.18  & <0.001  \\
Just right $|$ More AI      & 0.82        & 0.64  & 1.05  & 0.109   \\
AI Positive          & 0.37        & 0.27  & 0.52  & <0.001  \\
AI Dependency         & 0.05        & 0.04  & 0.08  & <0.001  \\
AI Contestability    & 0.08        & 0.05  & 0.11  & <0.001  \\
AI Assessability     & 0.05        & 0.03  & 0.07  & <0.001  \\
\midrule
Observations                & 1198        &       &       &         \\
R$^2$ Nagelkerke            & 0.324       &       &       &         \\
\bottomrule
\end{tabular}
\caption{Dependent variable is loss of control. Ordinary least squares regression model with 95\% confidence intervals.}
\end{table}

\subsection{Bail case}
\begin{table}[H]
\centering
\begin{tabular}[t]{lcccc}
\toprule
Predictors                  & Estimate & LL    & UL    & p       \\
\midrule
Intercept                   & 3.26     & 3.09  & 3.44  & <0.001  \\
AI Positive          & 0.38     & 0.14  & 0.63  & 0.002   \\
AI Dependency         & -0.57    & -0.83 & -0.32 & <0.001  \\
AI Contestability    & -0.41    & -0.66 & -0.16 & 0.001   \\
AI Assessability     & -0.51    & -0.75 & -0.27 & <0.001  \\
\midrule
Observations                & 1198     &       &       &         \\
R$^2$ / R$^2$ adjusted      & 0.065 / 0.062 &   &    &         \\
\bottomrule
\end{tabular}
\caption{Dependent variable is trust in government. Ordinary least squares regression model with 95\% confidence intervals.}
\end{table}

\begin{table}[H]
\centering
\begin{tabular}[t]{lcccc}
\toprule
Predictors                  & Estimate & LL    & UL    & p       \\
\midrule
Intercept                   & 4.21     & 4.04  & 4.38  & <0.001  \\
AI Positive          & 0.43     & 0.19  & 0.67  & 0.001   \\
AI Dependency         & 1.20     & 0.95  & 1.44  & <0.001  \\
AI Contestability    & 1.01     & 0.76  & 1.25  & <0.001  \\
AI Assessability     & 1.16     & 0.92  & 1.40  & <0.001  \\
\midrule
Observations                & 1198     &       &       &         \\
R$^2$ / R$^2$ adjusted      & 0.108 / 0.105 &   &    &         \\
\bottomrule
\end{tabular}
\caption{Dependent variable is loss of control. Ordinary least squares regression model with 95\% confidence intervals.}
\end{table}

\begin{table}[H]
\centering
\begin{tabular}[t]{lcccc}
\toprule
Predictors                  & Odds Ratios & LL    & UL    & p       \\
\midrule
Less AI $|$ Just right      & 0.34        & 0.26  & 0.43  & <0.001  \\
Just right $|$ More AI      & 2.10        & 1.64  & 2.68  & <0.001  \\
AI Positive          & 0.62        & 0.45  & 0.86  & 0.004   \\
AI Dependency         & 0.16        & 0.11  & 0.23  & <0.001  \\
AI Contestability    & 0.19        & 0.13  & 0.27  & <0.001  \\
AI Assessability     & 0.16        & 0.11  & 0.23  & <0.001  \\
\midrule
Observations                & 1198        &       &       &         \\
R$^2$ Nagelkerke            & 0.166       &       &       &         \\
\bottomrule
\end{tabular}
\caption{Dependent variable is loss of control. Ordinary least squares regression model with 95\% confidence intervals.}
\end{table}

\section{Pretest analysis with treatment checks}
\label{SI:pretest}
In this section, we report the results of the treatment checks we used in the pretest. For the pretest, we used a sample of 301 participants accessing the same online panel from Prolific that we used for the main study. These 301 participants from the pretest were not part of the sample we used for the main study later.

\begin{table}[H]
\begin{tabular}[t]{lcccc}
\toprule
Predictors                                 & Estimate & LL    & UL    & p       \\
\midrule
Intercept)                               & 3.27     & 2.96  & 3.59  & <0.001  \\
AI Positive                      & 0.81     & 0.51  & 1.10  & <0.001  \\
AI Dependency                      & 1.57     & 1.28  & 1.87  & <0.001  \\
AI Contestability         & 1.31     & 1.01  & 1.61  & <0.001  \\
AI Assessability                    & 0.84     & 0.54  & 1.14  & <0.001  \\
\midrule
\multicolumn{5}{l}{\textbf{Random Effects}}\\
$\sigma^2$                                 & 1.50     &       &       &         \\
$\tau_{00}$ participant                           & 1.41     &       &       &         \\
$\tau_{00}$ issue                          & 0.03     &       &       &         \\
ICC                                        & 0.49     &       &       &         \\
\midrule
N issue                                    & 3        &       &       &         \\
N participant                                     & 301      &       &       &         \\
Observations                               & 903      &       &       &         \\
Marginal $R^2$ / Conditional $R^2$         & 0.087 / 0.535 & &    &         \\
\bottomrule
\end{tabular}
\caption{Mixed‐effects model for the manipulation check variable dependency ("The government in this country could no longer function effectively without relying on AI systems") with 95\% confidence intervals.}
\end{table}

\begin{table}[H]
\centering
\begin{tabular}[t]{lcccc}
\toprule
Predictors                             & Estimate & LL    & UL    & p       \\
\midrule
Intercept                              & 4.18     & 3.93  & 4.43  & <0.001  \\
AI positive                            & 0.36     & 0.08  & 0.63  & 0.010   \\
AI Dependency                          & 0.83     & 0.55  & 1.10  & <0.001  \\
AI Contestability                      & 0.92     & 0.65  & 1.19  & <0.001  \\
AI Assessability                       & 1.00     & 0.73  & 1.28  & <0.001  \\
\midrule
\multicolumn{5}{l}{\textbf{Random Effects}}                                         \\
$\sigma^2$                              & 1.27     &       &       &         \\
$\tau_{00}$ participant                & 1.22     &       &       &         \\
$\tau_{00}$ issue                      & 0.01     &       &       &         \\
ICC                                     & 0.49     &       &       &         \\
\midrule
N issue                                 & 3        &       &       &         \\
N participant                           & 301      &       &       &         \\
Observations                            & 903      &       &       &         \\
Marginal $R^2$ / Conditional $R^2$      & 0.054 / 0.520 & &   &         \\
\bottomrule
\end{tabular}
\caption{Mixed‐effects model for the manipulation check variable transparency ("In this country, it is difficult to scrutinize how the government reaches its decisions") with 95\% confidence intervals.}
\end{table}

\begin{table}[H]
\centering
\begin{tabular}[t]{lcccc}
\toprule
Predictors                  & Estimate & LL     & UL     & p       \\
\midrule
Intercept                   & 3.84     & 3.60   & 4.09   & <0.001  \\
AI positive                 & -0.23    & -0.53  & 0.06   & 0.117   \\
AI Dependency               & -0.69    & -0.98  & -0.39  & <0.001  \\
AI Contestability           & -0.90    & -1.19  & -0.61  & <0.001  \\
AI Assessability            & -0.80    & -1.10  & -0.51  & <0.001  \\
\midrule
\multicolumn{5}{l}{\textbf{Random Effects}}                                             \\
$\sigma^2$                  & 1.51     &        &        &         \\
$\tau_{00}$ participant     & 1.04     &        &        &         \\
$\tau_{00}$ issue           & 0.00     &        &        &         \\
ICC                         & 0.41     &        &        &         \\
\midrule
N issue                     & 3        &        &        &         \\
N participant               & 301      &        &        &         \\
Observations                & 903      &        &        &         \\
Marginal $R^2$ / Conditional $R^2$ & 0.044 / 0.435 & & &         \\
\bottomrule
\end{tabular}
\caption{Mixed‐effects model for the manipulation check variable contestability ("If I feel that I have been treated unfairly by the state, I have the opportunity to effectively challenge a state decision") with 95\% confidence intervals.}
\end{table}

\section{Ethics}

The research protocol has been submitted for review to the Institutional Review Board (IRB) at the home institution of one of the authors. The experiment involved no deception: all vignette information was explicitly framed as hypothetical, and respondents received no false or misleading statements at any point. Participation was voluntary, informed consent was obtained before the survey, and confidentiality was guaranteed throughout. To ensure full transparency, every respondent saw a debrief screen immediately after completing the questionnaire that reiterated the hypothetical nature of all scenarios and provided contact details for further questions. Below we provide a structured ethics appendix in line with the suggestion provided by \citeASI{asiedu_call_2021}.

\begingroup
\small                   
\setlength\LTleft{0pt}
\setlength\LTright{0pt}
\begin{longtable}{@{}p{0.25\textwidth}p{0.70\textwidth}@{}}
\caption{Ethics Appendix}
\label{tab:ethics}\\
\toprule
\textbf{Topic} & \textbf{Discussion} \\
\midrule
\endfirsthead


\bottomrule
\endfoot
    Policy equipoise \& scarcity 
      & The factorial survey experiment compares alternative vignette descriptions; no respondent is denied an otherwise available public benefit or exposed to material risk. Each arm offers equal—and minimal—psychological stakes. Hence policy equipoise obtains: there is no expert consensus that any vignette condition is better or worse for respondents, and no scarce resource is being withheld. Randomisation therefore raises no distributive-justice concern. \\[0.75em]
    Role of the researcher 
      & Researchers conceived the theoretical framework, wrote the vignettes, and fielded the online survey through Prolific. They did not exercise decision‐making power over actual AI implementation in government. \\[0.75em]
    Potential harms to participants 
      & Participation involves reading short policy scenarios and answering attitudinal questions. Possible harms are limited to mild fatigue or transient concern about AI. No identifying information beyond platform IDs is collected, limiting potential harms due to data-privacy violations. \\[0.75em]
    Potential harms to non-participants 
      & The survey introduces no intervention in the real world; therefore it poses no direct risk to non-participants. \\[0.75em]
    Conflicts of interest 
      & The authors declare no financial, professional, or personal interests that could reasonably be perceived to bias the study. Funding (internal university seed grant) was unconditional and carried no publication or disclosure restrictions. \\[0.75em]
    Intellectual freedom 
      & All hypotheses were pre-registered, and the full analysis script is archived on OSF. Funders exerted no control over design or analysis. Authors retain the right to publish regardless of outcome and commit to open sharing of de-identified data and code upon journal acceptance. \\[0.75em]
    Feedback to participants 
      & Participants were debriefed with information on the study’s goals and objectives. \\[0.75em]
    Foreseeable misuse of results 
      & Findings could be cited to oppose any use of AI in government or, conversely, to downplay legitimate control concerns. We ensured our discussion represents a balanced view of the findings. \\
\end{longtable}
\endgroup

\clearpage
\phantomsection
\bibliographystyleSI{apacite}
\bibliographySI{principal_agent_SI}
\end{document}